\documentstyle[preprint,prb,aps,amssym,epsf]{revtex}
\newcommand{\bmp}{{\mbox{\boldmath $p$}}}

\newcommand{\bmr}{{\mbox{\boldmath $r$}}}

\newcommand{\bmi}{{\mbox{\boldmath $i$}}}

\newcommand{\AAr}{{\rm \AA}}
\newcommand{\Ain}{{\rm {\AA}^{-1}}}

\begin{document}
\preprint {WIS-10/00 June-DPP}
\draft
 
\date{\today}
\title{A novel method for the extraction of the condensate fraction
of liquid $^4$He}
\author{A.S. Rinat and M.F. Taragin}
\address{Department of Particle Physics, Weizmann Institute of Science,
         Rehovot 7616, Israel}
\maketitle
\begin{abstract}
 
We present a new, semi-phenomenological method to extract the  condensate
fraction $n_0(T)$  from two sets of measured responses of $^4$He with
identical kinematics at $T<T_c$ and  $T>T_c$.
 
\end{abstract}
 
\section{Introduction}
 
Over the past decades attempts have been made to determine the condensate
fraction $n_0(T_<)$ of liquid $^4$He, which is
defined as the asymptotic limit of
the single-atom density matrix (SADM) $\rho_1(0,s;T_<)/\rho(T_<)$.
\begin{eqnarray}
n_0(T_<)=\lim_{s \to \infty}\rho_1(0,s;T_<)/\rho(T_<)
\label{a1}
\end{eqnarray}
$\rho(T)=\rho_1(0,0;T)$ is the number density. We shall use the notation
 
$$T_<=T<T_c;T_>=T>T_c;T_{\gtrless}=T_<,T_>$$
 
In  spite of  the increasing  quality of  data, the  various experimental
results show an  unsatisfactory spread in $n_0$.  The reason  may well be
that there is no unequivocal, direct link between the above definition of
$n_0(T_<)$  and experimental  information.  The  quality of  an extracted
condensate fraction thus depends on the  accuracy of the approximate
isolation of the relevant SADM, which contains the condensate fraction.
 
Without striving for completeness we briefly review:
 
i) computations from first principles;
 
ii) results from the comparison of specific theories and data;
 
iii) data analysis with minimal theoretical input;
 
We then propose and evaluate a  novel extraction method which, except for
minor assumptions, is largely model-independent.

$Computations$:  In principle  stochastic  methods  for a  representative
sample  of a  finite number  of atoms  determine the  ground state  wave
functions and  thus $\rho_1$.  Those  wave functions, and thus  the SADM,
can be  computed up to large  values of $s$ and  (\ref{a1}) thus directly
provide $n_0$ (see for instance [\onlinecite{whit,mor}]). In practice the
method is accurate only for $T=0$  K, and from sufficiently large $s$ one
extracts $n_0(T=0)$.   Nevertheless, even for $T=0$, different stochastic
methods produce a spread in $n_0$ of, as much as 25\% \cite{mor}.
 
For finite $T$ a calculation of $\rho_1$ requires a canonical average
\begin{eqnarray}
\rho_1(0,s;T)/\rho(T)=Z^{-1}(\beta)\int d\bmr_1 \bigg (
\Pi_{k\ge 2}\int d\bmr_k\bigg )
\langle\bmr_1,\bmr_k|{\rm exp}[-\beta H_A]|\bmr_1-s\hat\bmi{_z},\bmr_k)
\rangle
\label{a2}
\end{eqnarray}
with $Z(\beta)$, the partition function.
 
Results  for $\rho_1(0,s;T\ne  0$) known  to us,  are for  $s \lessapprox
7\AAr$  and  for  a  number of  discrete  $T_\gtrless$  around  $T_c\,\,$
\cite{cep}, but  beyond $s\approx  2.5\AAr$ the computed  SADM apparently
have sizable inaccuracies.  This  reflects on condensate fractions, which
have  been extracted  from $s$  in the  range $\approx  4-7\AAr$ where  a
constant value of the SADM is only reached on the average \cite{cep}.
 
$Theory$: Efforts to obtain the condensate fraction naturally focuss on
the extraction of the SADM from an observable.  A prime example  is the
cross  section  for   inclusive  neutron  scattering  on   He,  which  is
proportional to  the response $\phi(q,y)$.   The latter is a  function of
two kinematic  variables, e.g. the momentum transfer $q$  and the
GRS-West scaling  variable, $y=(M/q)(\nu-q^2/2M)$, with $M$,  the mass of
an atom.  For fixed $q$ the variable $y$ is a measure for the energy loss
$\nu$ \cite{grs1,west}.
 
In order to  compare with experiment the above response  has to be folded
into the  experimental resolution  function $E$.   One may  then formally
write the response and its Fourier Transform (FT)
in a closed  form (see for  instance Ref. [\onlinecite{grs1,grs2}])
\begin{mathletters}
\label{a3}
\begin{eqnarray}
\phi^{th,conv}(q,y;T)
&=&\int dy'\int dy''F_0(y';T)R(q,y-y';T)E(q,y'-y'';T)
\label{a3a}\\
\tilde\phi^{th,conv}(q,s;T)&=&\int dy\,{\rm exp}^{-isy}
\phi^{th,conv}(q,y;T)=\tilde F_0(s;T)\tilde R(q,s;T)\tilde E(q,s;T)
\label{a3b}
\end{eqnarray}
\end{mathletters}
The first factor $F_0(y;T)$ is the  asymptotic limit of the response, and
depends  on  the  single-atom  momentum  distribution  $n(\bmp,T)$.   Its
Fourier Transform is  just the SADM.  Final State  Interactions (FSI) are
formally contained in  $R$, which can be expanded in  powers of $1/q$ and
which depends on higher order  density matrices $\rho_n$.  In practice it
suffices to retain only the dominant FSI $\propto \rho_2/q$. Finally, $E$
is Experimental Resolution (ER) of the measuring devise.
 
In the standard  approach one computes from dynamics the  FSI factor $R$,
which  modifies the  asymptotic response  $F_0$ for  finite $q$  (see for
instance    Ref.  [\onlinecite{grs1,grs2,rt}]). In addition many-body
density   matrices  which weigh  all components are required.
We  focus on  $F_0$,  which
is related to  the  momentum
distribution $n(\bmp;T)$ and which for $T_<$ requires modeling
\begin{eqnarray}
n(\bmp,T_<)&=&n_0(T_<)[(2\pi)^3\delta(\bmp)+f(p;T_<)]+A(T_<)n^{no}(p;T_<)
\nonumber\\
A(T_<)&=&1-n_0(T_<)[1+\tilde f(0;T_<)]
\label{a4}
\end{eqnarray}
In  the above  parametrization (\ref{a4})  one finds  in addition  to the
macroscopic  fraction   $n_0(T_<)$  of  atoms  with   momentum  $\bmp=0$,
$f(p,T_<)$,  the  fraction  of  atoms  relative  to  $n_0$  with  momenta
$p\lesssim  p_c$  in  the  immediate neighborhood  of  $p=0$  \cite{gav}.
Finally,  $n^{no}(p,T)$  above  is  the  normal  part  of  the  momentum
distribution  of  atoms with  $p\gtrsim  p_c$;  $A(T)$ cares  for  proper
normalization.
 
Albeit dominant  for $p\ne 0,\,n^{no}(p,T_<)$ is  not directly measurable
and  is   for  low   $T$  frequently   assumed  to   be  $T$-independent:
$n_0^{no}(\bmp;T_<)\leftrightarrow  n_0^{no}(\bmp;T_>)=n(\bmp;T_>)$.  For
$T_>$ it is the measurable $n(\bmp;T)$.  Equivalently for the SADM
\begin{mathletters}
\label{a5}
\begin{eqnarray}
\rho_1(0,s;T_<)/\rho(T_<)&=&n_0(T_<)+G(s;T_<)
\label{a5a}\\
G(s;T_<)&=&n_0(T_<)\tilde f(s;T_<)+A(T_<)[1+\tilde f(0,T_<)]
\rho_1^{no}(s;T_<)/\rho(T_<)
\label{a5b}
\end{eqnarray}
\end{mathletters}
Above, $G$ is the difference  between the  SADM and its  asymptotic limit
$n_0$,  which  will  be  called  the  deficit  function.   By  definition
${\lim_{s \to \infty}}G(s,T_<)=0$.
 
We summarize: Except  for $T=0$, there is no accurate  information on the
SADM  for  large  $s$.   For  finite $s$  an  expression  for  $G$,  e.g.
(\ref{a5b}), is model-dependent and without  knowledge of
both $\rho_1(0,s;T_<)$ and  $n_0(T_<)$ there is no accurate  way to reach
the deficit function $G$.
 
$Data-analysis$: One method used in the  past relates the integral of the
difference between  responses for $T=T_\gtrless$ in  the region $y\approx
0$ to the fraction of atoms having momenta in that region.  The method is
not accurate because of imprecise isolation of FSI and ER broadening
in (\ref{a3}) (see for  instance Ref. [\onlinecite{sss1}]).
 
Of similar  nature are  cumulant parametrizations  of reponses  for $T_<$
\cite{glyde}.  Even if  the above assumption on  $n^{no}(p,T_<)$ is made,
the data do not permit the extraction of an accurate value of $n_0(T_<)$
and $p_c$. Instead one studies fits for varying $n_0(T_<)$
\cite{glyde,azu}.
 
\section{Consistency relations and  a novel method for  the extraction of
the condensate fraction.}
 
We  now describe  a  new  method for  the  extraction  of the  condensate
fraction  as it  appears in  the SADM  for $T_<$.   We try  to avoid  the
described insufficient  information on the  latter by exploiting  data on
inclusive  scattering  on  $two\,\,T_{\gtrless}$ at  precisely  the  same
momentum  transfer $q$  and energy  loss  $\nu$, or  the related  scaling
variable $y$.   Fairly recent data of  that type exist for  $T=1.6,2.3$ K
\cite{az3,foot} (MARI data).
 
We  first define  from  Eq. (\ref{a3b})  the  ratio of  the  FT of  those
corresponding data
\begin{mathletters}
\label{a6}
\begin{eqnarray}
\eta^{exp}(q,s;T_<,T_>)&\equiv &\bigg\lbrack \frac{\tilde\phi_E(q,s;T_>)}
{\tilde\phi_E(q,s;T_<)}\bigg\rbrack^{conv}
=\bigg \lbrack \frac{\tilde\phi(q,s;T_>)}{\tilde\phi(q,s;T_<)}
\bigg\rbrack^{deconv}
\label{a6a}\\
&\approx&\bigg\lbrack
\frac{\rho_1(0,s;T_>)/\rho(T_>)}{\rho_1(0,s;T_<)/\rho(T_<)}\bigg\rbrack
\label{a6b}
\end{eqnarray}
\end{mathletters}
Eq. (\ref{a6a})  expresses the  $T$-independence of the  ER for  the MARI
data  and as  a  consequence  the ratio  of  actual  convoluted data  are
replaced by deconvoluted ones.  Those  $\eta^{exp}$ depend on $q$ and are
complex, because the FT of the data are complex.
 
Eq.  (\ref{a6b}) is  of  a different  nature.  It  assumes  that the  FSI
function $R$ is at most  weakly $T$- dependent.  Although verified within
the accuracy of the data  \cite{glyde,azu}, we cite weak $T$-dependence
on the measured pair-distribution function $g(r,T)$ \cite{sven} which, at
least in some theories, enters the FSI function $R\,\,$ \cite{grs1}.  The
price for $'$removing$'\,R$ and the ER is the appearance of
two SADM, instead of the desired $\rho_1(0,s;T_<)/\rho(T_<)$.
 
Since the precision with which one ultimately extracts the condensate
fraction depends on the  quality of the input, we discuss the  latter. We
note that
$\eta^{exp}$ in (\ref{a6})  uses not actual data, but  instead their FT.
With substantial  noise, in particular in  the larger $ y $ tails  of the
data for  $\phi(q,y)$, one  cannot avoid smoothing,  or even  cutting out
those low-intensity data, which affect the accuracy of $\eta^{exp}$.
 
We  first  apply  the  method  to   the  above-mentioned  MARI  data  at
$T_{<}=1.6\,{\rm  K};  T_>=2.3$  K  \cite{az3}  from  which  we  chose  a
representative sample of highest quality for which $q=17,21,23,29  \Ain$.
From those we extract    the   ratio   ${\rm   Im}[\eta^{exp}(q,s)]/{\rm
Re}[\eta^{exp}(q,s)]$, which  is well determined in  the range $s\lesssim
2.5 \AAr$.  For growing $s$, the imaginary part increases  from 0 but is,
even at medium
$s\approx 2.5 \AAr$, only 2\% of the real part,  and can thus for
all purposes be neglected.  This confirms  that at least the ratio of the
involved FSI functions $R$ is virtually $T$-independent.
 
Next we investigate the $q$-dependence of ${\rm Re}[\eta^{exp}]$ and show
in Fig.  1 data for some  individual $q$.  Before judging  the quality we
mention that,  contrary to predictions,  some data for $T_>$  and varying
$q$, and more for $T_<$, are not uniformly smooth in $q$.  This may cause
$apparent\,\, q$-dependence in $\eta$ and justifies the use of an average
for $s\lesssim 2.5 \AAr$
\begin{eqnarray}
\eta^{exp,av}(s;T_{\gtrless})
&\equiv& \bigg\langle \eta^{exp}(q,s;T_{\gtrless})\rangle_q
\approx \bigg\langle{\rm Re} [\eta^{exp}(q,s;T_{\gtrless})]\bigg\rangle_q
\label{a7}
\end{eqnarray}
 
\vskip1cm
\begin{minipage}{13cm}
\begin{center}
\leavevmode
\epsfxsize=15cm
\epsffile{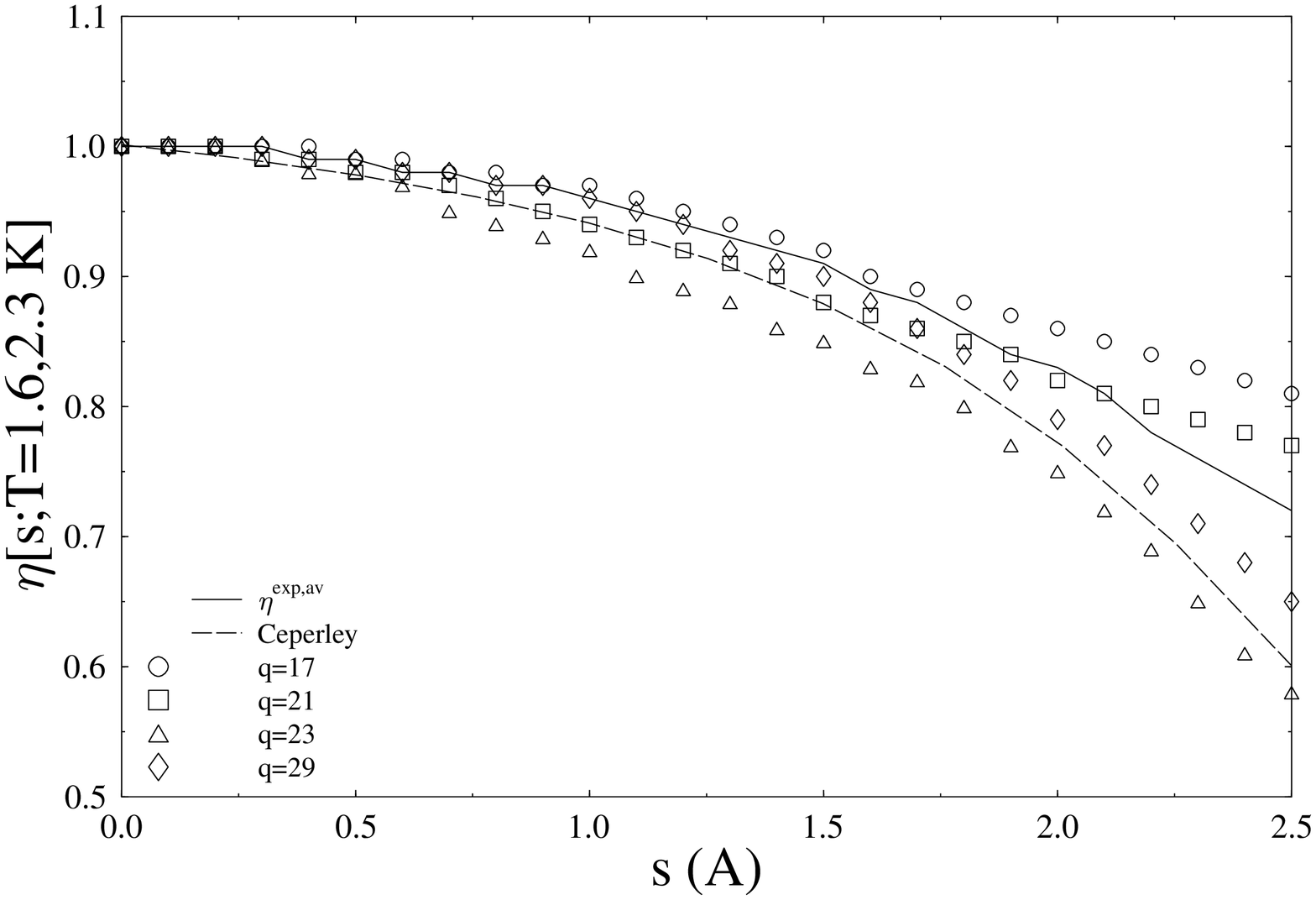}
\end{center}
{\begin{small}
 
Fig.  1.   The  ratio  $\eta^{exp}(s;  T_<=1.6{\rm  K},T_>=2.3$  K,)  Eq.
(\ref{a5a}),  for  $q=17,  21,  23,  29  \Ain$  and  its  average
(\ref{a6}) over  $q$ (drawn line).   The long dashes  give $\eta^{comp}$,
Eq.  (\ref{a8}).
 
\end{small}}
\end{minipage} \\ \\

 
As a first step we test (\ref{a6b}) with the only   available   computed
results  on  SADM   for $T_{\gtrless}$ \cite{cep}. Without any dynamical
cause, the outcome for $T$=1.54 K closest to $T$=1.6 K of the MARI data,
is not very close to values from interpolation between neighboring $T$.
We suspect
the mentioned inaccuracies which  are  also  apparent in  non-vanishing
condensate  fractions   for  $T_>$  and  the   non-smooth  dependence  of
$n_0(T_<)$ as function  of $T$ outside their  estimated theoretical error
bars.  Nevertheless we use interpolated  computed results and  define for
$T_<=1.6\,{\rm K};T_>=2.3$ K
\begin{eqnarray}
\eta^{comp}(s;1.6,2.3)&\equiv&
\bigg\lbrack \frac{\rho_1(0,s;2.3)/\rho(2.3)}{\rho_1(0,s;1.6)/\rho(1.6}
\bigg\rbrack^{comp}
\label{a8}
\end{eqnarray}
The above as function of $s\leq 2.5 \AAr$ is also entered
in Fig. 1 and is
seen to represent the data rather  poorly.  We shall thus exclusively use
$\eta^{exp,av}$  and  return  to  Eq.  (\ref{a5a}),  which  by  means  of
(\ref{a7}) and (\ref{a6b}) becomes
\begin{eqnarray}
n_0(T_<)=\frac {\rho_1(0,s;T_>)/\rho(T_>)}{\eta^{exp,av}(s,T_{\gtrless})}
-G(s;T_<)
\label{a9}
\end{eqnarray}
We note that, contrary  to (\ref{a1}),  Eq.  (\ref{a9}) is not applicable
in the  asymptotic
region, because in order to produce a constant, numerator and denominator
in the ratio in (\ref{a9}) have to tend to zero in exactly the same
fashion. This is an impossibly stringent demand.
 
Although Eq. (\ref{a9}) provides in principle a value of the condensate
fraction for $arbitrary,\,\, finite\,s$, it is clearly  of greater
interest to  consider a range  of $s$.  In fact,  the right hand  side of
(\ref{a9})  is  a function  of  $s$,  whereas the  left  hand  side is  a
constant,  implying a consistency  test, to  be  passed as  a
condition  for a  succesful extraction  of $n_0(T_<)$.   The size  of the
$s$-range depends on  the available information on  the various functions
in (\ref{a9}).
 
The  most delicate  source of  information is  the deficit  function $G$.
After the  warnings above, one clearly  does not want to  use (\ref{a5b})
and has thus  to rely on the model-independent, not  too accurate results
of    Ref.    [\onlinecite{cep}],    which    for    $4\lesssim    s({\rm
in}\,\AAr)\lesssim  7$, provide  $n_0(T)$ with  uncertainties of  the
order of  15\%.  The crucial  point is that those  uncertainties decrease
relatively to  $G(s;T_<)$ which  increases for  decreasing $s$.   We thus
conclude that, up  to medium $s\lesssim 2.5 \AAr$, one  can trust and use
the computed $G$.  The  above happens to also be the  range for which the
available data determine $\eta$ sufficiently well.
 
We thus perform the consistency test,  implied by (\ref{a8}) and it comes
as a  surprise that the  right hand side  in the considered  $s$-range is
only weakly $s$-dependent, leading to
\begin{eqnarray}
n_0(1.6)=0.0625\pm 0.0017
\nonumber\\
n_0(0)= 0.090\pm 0.030
\label{a10}
\end{eqnarray}
The latter  is the  result for  a standard  extrapolation to  $T=0$. The
small error  limits are due  to averaging  results for different  $s$; we
could not estimate the same for uncertainties in $G$.
We  note  that the  errors  on  the  three  functions in  (\ref{a9})  are
uncorrelated, which  underscores the feducity  of the extracted  value of
$n_0(T_<)$.
 
The above numbers use $G(s;T=1.6 {\rm K})$, interpolated between values
for $T$= 1.54,1.82K as reported by Ceperley. We mentioned that the former
set, does not interpolate smoothly between values for $T$=1.18, 1.82 K.
Smoothing leads to a lower condensate fraction. Obviously the accuracy
with which one can extract the condensate fraction depends on the
same for the input.
 
Above we  also listed older Argonne  data for fixed $q=23.1  \Ain$ and
some 10 temperatures around $T_c$.  Unfortunately we do not have
available the actual data and the ER functions, and have to analyze
instead parametrizations of the above.
as given by the authors. Those are given by the authors in the form of
double  Gaussians for data deconvoluted from  ER,
and  in addition also deconvoluted from  FSI.  The
information  no doubt  reduces the  required  accuracy. Yet,  in view  of
scarcity of information we performed the above analysis for the Argonne
data. In order to compare with the MARI data and we fix $T_>$ at 2.3 K.
 
One first observes that the computed
FT of the above parametrized data are essentially real.  Next,
ratios of those FT for and $T_<$ and the above $T_>$ are identical within
1\%, in support of (\ref{a6}). Nevertheless, only a limited part of  the
information can be used for an analysis of the above type. Thus for
$T_c\ge T\gtrsim 1.8$ K, $n_0(T_<)$ is the difference of nearly equal
terms in (\ref{a9}). A reliable determination
requires  a  precision on  $\rho_1(0,s;T_>)/\rho(T_>)$
and  $G(s;T_<)$, which is beyond the Ceperley results.
A different difficulty occurs for $T\lesssim 1$ K, where  one has to
make an uncertain extrapolation,  using the  lowest  $T$ results  of
Ceperley.  The following outcome
\begin{eqnarray}
n_0(T=1.0\, {\rm K})=0.063 \pm 0.006
\nonumber\\
n_0(T=1.5\, {\rm K})=0.060 \pm 0.008
\nonumber\\
n_0(T=1.8\, {\rm K})=0.050 \pm 0.005
\label{a11}
\end{eqnarray}
carries some 20-25\%  uncertainties, which appreciably exceed  the one in
(\ref{a10}) based on the MARI data.  One notes that $n_0(T\approx 1.6$ K)
from the MARI and Argonne data  approximately agree.  However, in view of
the fact that we  had to use parametrizations of the  Argonne data, we do
not attach  too much  significance to  the correct  trend of  $n_0(T)$ as
function  of $T$  and the  somewhat  low, but  not unreasonable,  average
$\langle n_0(T=0)\rangle_{T_<}$=0.079.
 
In conclusion, we  have suggested and worked out a  new method to extract
the condensate  fraction in  $^4$He from the  Fourier transforms  of data
sets on  structure functions at  the same momentum and  energy transform,
but for two  $T$ below and above $T_c$.  In  addition the method requires
some   previously  computed   dynamical   information.   The   expression
(\ref{a9}) for $n_0(T_<)$ appears in principle  as a function of $s$, but
is  in practice  a  well-determined and  thus  meaningful constant,  the
condensate fraction.

 
{Figure Captions}


\begin{references}
 
\bibitem{whit}
P.A. Whitlock and R.M. Panoff, Can. J. Phys. 65, 1409 (1987);
C. Carraro and S.E. Koonin, Phys. Rev. Lett.  65, 2792 (1990).
J. Boronat and J. Casulleras, Phys. Rev. B 49, 8920 (1994).
 
\bibitem{mor}
S. Moroni, G. Senatore and S. Fantoni, Phys. Rev. B 55, 1040 (1997).
 
\bibitem{cep}
D.M. Ceperley and E.L. Pollock, Can. J. Phys. 65, 1416 (1987) and
private communication (1989).
 
\bibitem{grs1}
H.A. Gersch, L.J. Rodriguez and Phil N. Smith, Phys. Rev. A 5,
1547 (1973).
 
\bibitem{west}
G.B.  West, Phys. Reports C 18, 264 (1975).
 
\bibitem{grs2}
H.A. Gersch and L.J. Rodriguez, Phys. Rev. A 8, 905 (1993).
 
\bibitem{rt}
A.S. Rinat and M.F. Taragin,  Phys. Rev. B 58, 15011 (1998).
 
\bibitem{gav}
J. Gavoret and P. Nozieres, Ann. of Phys. (N.Y.) 28, 349 (1974).
 
\bibitem{sss1}
P.E Sokol, T.R. Sosnick and W.M. Snow, in 'Momentum Distributions', ed.
by Richard N. Silver and Paul E. Sokol, Plenum Press, NY and London,
p. 139, 1988.
 
\bibitem{glyde}
H.R. Glyde, Phys. Rev. B 50, 6726 (1994).
 
\bibitem{azu}
R.T. Azuah, W.G. Stirling, H.R. Glyde, P.E. Sokol and S.M. Bennington,
Phys. Rev. B 56, 14620 (1997).
 
\bibitem{az3}
R.T. Azuah, PhD Thesis, Univ. of Keele, UK (1994).
 
\bibitem{foot}
T.R. Sosnick, W.M. Snow and P.E. Sokol, Phys. Rev. B 41, 11185 (1990).
 
\bibitem{sven}
E.C. Svensson, V.F. Sears, A.D.B. Woods and P. Martel, Phys. Rev.
B 21, 3638 (1980).
 
\end{references}
\end{document}